\documentclass[aps,prb,superscriptaddress,amsmath,amssymb,twocolumn,floatfix,reprint,longbibliography]{revtex4-2}
\usepackage{graphicx}
\usepackage{bm}
\usepackage{times}
\usepackage[colorlinks=true,citecolor=blue,linkcolor=blue,urlcolor=blue]{hyperref}
\graphicspath{{figs/}}
\newcommand{\bk}{\bm k}
\newcommand{\ii}{\mathrm{i}}
\newcommand{\dd}{\mathrm{d}}
\newcommand{\Tr}{\operatorname{Tr}}
\AddToHook{env/thebibliography/begin}{\interlinepenalty=10000\relax}
\ExplSyntaxOn
\DeclareDocumentCommand{\bibfield}{mm}
  {
    \str_if_eq:nnTF {#1} {journal}
      {
        \tl_set:Nn \l_tmpa_tl {#2}
        \tl_replace_all:Nnn \l_tmpa_tl {\ ,} {,}
        \tl_use:N \l_tmpa_tl
      }
      {#2}
  }
\ExplSyntaxOff

\begin{document}
\title{Fractional Chern Insulator in a Finite-Range One-Band Model}
\author{Tong Liu}
\affiliation{Department of Physics, Hong Kong University of Science and Technology, Clear Water Bay, Hong Kong, China}
\affiliation{Center for Theoretical Condensed Matter Physics, Hong Kong University of Science and Technology, Clear Water Bay, Hong Kong, China}
\author{Hoi Chun Po}
\affiliation{Department of Physics, Hong Kong University of Science and Technology, Clear Water Bay, Hong Kong, China}
\affiliation{Center for Theoretical Condensed Matter Physics, Hong Kong University of Science and Technology, Clear Water Bay, Hong Kong, China}
\begin{abstract}
A common route for realizing fractional Chern insulators (FCIs) is to engineer bands that mimic the physics of Landau levels, which serve as the single-particle topological background for the emergence of strongly correlated fractional quantum Hall states.
Here we show
that FCIs can also arise in a one-band model devoid of any single-particle quantum geometry.
Through a systematic trivialization of the single-particle quantum geometry in FCI models that go beyond the Landau-level paradigm, we construct a single-band, interaction-only model on the square lattice, with all terms confined to individual plaquettes. At one-third filling, exact diagonalization and
density-matrix renormalization-group calculations reveal a three-state
low-energy manifold, a Hall conductance of $-2e^2/(3h)$, and fractional
charge pumping without translation-symmetry breaking. We trace this
behavior to nontrivial topology in the repulsive two-particle sector,
offering a microscopic perspective on the strongly correlated origin of
fractional Hall order.
\end{abstract}
\maketitle

\section{Introduction}
Fractional Chern insulators (FCIs) are lattice analogs of fractional
quantum Hall states without external magnetic
fields~\cite{Neupert2011,Sheng2011,Regnault2011,WangBoson2011,Wu2012Zoology,Wang2012NonAbelian,Liu2012HigherC}.
Recent experiments have realized FCIs in twisted bilayer
MoTe$_2$~\cite{Cai2023,Zeng2023,Park2023,Xu2023}
and rhombohedral graphene/hBN moir\'e
superlattices~\cite{Lu2024,Xie2025}.
These observations were preceded by more than a decade of theoretical
work, largely focused on interacting particles in isolated, nearly flat
Chern bands which mimic Landau levels (LLs)~\cite{Tang2011,Sun2011,Xiao2011}.
In this LL paradigm, the Bloch-state wavefunctions are optimized so that their quantum geometry approximates that of the LL wavefunctions~\cite{Claassen2015,Lee2017,Ledwith2020,Wang2021,Mera2021Engineering,Ledwith2022,Wang2022Hierarchy,Dong2023,Ledwith2023Vortex,Estienne2023,Liu2025Generalized}. The desired features include a nearly uniform Berry curvature, saturation of the quantum-metric trace bound~\cite{Roy2014,Jackson2015,Mera2021Kahler,Ozawa2021}, and projected density
operators that approximate the Girvin--MacDonald--Platzman (GMP)
algebra~\cite{Girvin1986,Parameswaran2012,Wang2023Ideal}.

Limiting to the LL paradigm restricts the search space for real-material candidates for FCIs. It is also of fundamental interest to understand how FCIs could be qualitatively different from the fractional quantum Hall states.
Ongoing efforts have therefore focused on discovering FCIs beyond the LL paradigm, say in bands with non-Landau-level continuum limits~\cite{Bauer2022} or far-from-ideal quantum geometry~\cite{Shavit2024,Fonseca2025}.
FCIs have also
been found in gapless flat bands~\cite{Yang2025}, even without a
well-defined Berry curvature at the band-touching point~\cite{Yang2026}.
In another direction, multi-band systems could develop nontrivial band topology at the mean-field level, which could in turn facilitate the emergence of FCIs~\cite{Simon2015,Zhou2024Moireless,Dong2024HallCrystal}. 

Single-particle quantum geometry, characterized by a nontrivial Chern number or a nearly ideal quantum metric, is a common ingredient that enables FCIs. In this work, we show that an FCI can also be stabilized in a one-band model using only finite-range interactions. With only one spinless orbital per primitive cell, the Bloch states are always strictly atomic provided translation symmetry remains unbroken. 
More concretely, our constructed model is defined on the square lattice, does not include any fermion hopping, and contains only quartic, time-reversal-breaking interaction terms within a plaquette. 
Curiously, we found a ground-state Hall conductance of $\sigma_{xy} = -(2/3) e^2/h$ at one-third filling of the atomic band, which deviates from the usual Laughlin-state expectations.
Such an interaction-only, quantum-geometry-free model could be viewed as a minimal setting for investigating the strongly correlated origins of FCIs beyond the LL paradigm.

We note that FCIs in a single-band model have also been constructed starting from the conformal field theory perspective, but the associated parent Hamiltonians are generally long-ranged~\cite{Tu2014LatticeLaughlin,Glasser2015LatticeMooreRead}. 
A recent work has investigated instead if exact FCI ground states could emerge in suitably designed, LL-inspired multi-band systems using only Hubbard interactions \cite{HerzogArbeitman2026ExactFCI}. Here, we address the opposite question and explore if an FCI could be stabilized without any underlying single-particle quantum geometry. Our construction is guided by a systematic trivialization of the single-particle quantum geometry, with a single atomic band as the endpoint. This approach is motivated by earlier work on FCIs in dice-lattice models. Ref.~\cite{Yang2025} found FCIs in a gapless flat band with nearly uniform Berry curvature away from the band-touching point.
Ref.~\cite{Lin2026} subsequently showed that hopping anisotropy can isolate a flat band with zero Chern number by breaking lattice symmetry, while still preserving the FCI at two-thirds filling. The band therefore loses its net Berry flux but retains the nontrivial local quantum geometry that stabilizes the FCI.
 Here, we instead start from the fine-tuned, $p4$-symmetric square-lattice
critical topological flat-band (CTFB) model of Ref.~\cite{Li2026}.
We first show that hybridizing the auxiliary \(s\) orbital with the physical \(p\) orbitals yields an isolated, exactly flat band with \(C=0\) while preserving \(p4\) symmetry. At one-third filling, repulsive interactions stabilize an FCI in this band.

Next, we construct the one-band model by Wannier downfolding the interactions onto the trivial flat band, followed by a finite-range truncation of the interactions and parameter tuning to stabilize the many-body gap.
This procedure leads to a one-orbital square-lattice Hamiltonian with no
single-particle hopping and all terms confined to an elementary
plaquette. We call this interaction-only Hamiltonian the plaquette
model; ED and fractional charge pumping in density-matrix
renormalization-group (DMRG) calculations establish its fractional
Hall state.
Furthermore, we show that although the single-particle quantum geometry has been removed after reduction to a one-band model, nontrivial topology in the two-particle channel remains. Our results pave the way for further investigating the correlated processes leading to FCIs beyond the LL paradigm.

\section{Three-band model}
\begin{figure}[!t]
 \centering
 \includegraphics[width=\columnwidth]{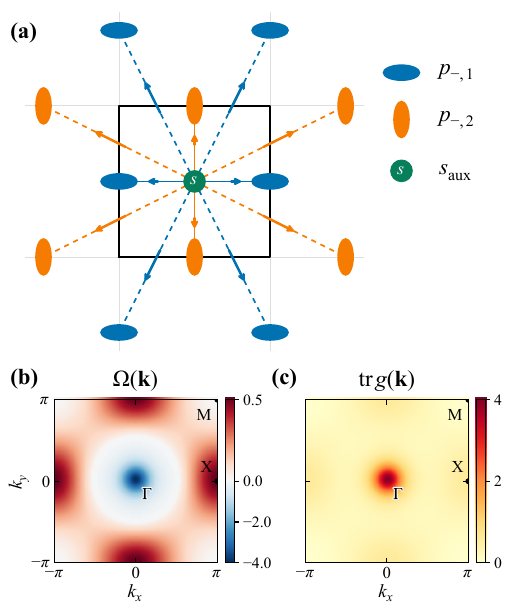}
 \caption{Three-band construction of an isolated trivial flat band.
 (a) Orbital arrangement inherited from the critical model of
 Ref.~\cite{Li2026}: $p_1$, $p_2$, and $s$ are blue, orange, and
 green. Lines indicate the underlying critical model's hopping
 geometry.
 (b),(c) Berry curvature and trace of the quantum metric, respectively,
 calculated from the Bloch state in Eq.~\eqref{eq:regularized-section} at \(\delta=1\).
 $\Gamma=(0,0)$, $X=(\pi,0)$, and $M=(\pi,\pi)$.}
 \label{fig:liyq-model}
\end{figure}

\begin{samepage}
We consider a square-lattice model whose exactly flat band has
the normalized Bloch wavefunction
\begin{equation}
 |\psi_\delta(\bk)\rangle=\frac{u_\delta(\bk)}{\sqrt{N_\delta(\bk)}},
 \qquad
 u_\delta(\bk)=
 \begin{pmatrix}
   F_1(\bk)\\
   F_2(\bk)\\
   \delta f(\bk)
 \end{pmatrix},
 \label{eq:regularized-section}
\end{equation}
where $\delta\in\mathbb R$, $N_\delta=u_\delta^\dagger u_\delta$,
$f=\cos(k_x/2)\cos(k_y/2)$. The first two components are
$F_1=-\cos(k_y/2)\sin k_x-2\ii\sin(k_y/2)$ and
$F_2=2\sin(k_x/2)+\ii\cos(k_x/2)\sin k_y$.
\par
\end{samepage}
The components refer to two spinless
$p_-=(p_x-\ii p_y)/\sqrt2$ orbitals, $p_1,p_2$, at the $2c$ Wyckoff positions
$\bm\tau_1=(0,1/2)$ and $\bm\tau_2=(1/2,0)$, and an auxiliary
$s$ orbital at the $1b$ position $\bm\tau_s=(1/2,1/2)$
[Fig.~\ref{fig:liyq-model}(a)].
We use the convention
$c_{\bk\alpha}=N_c^{-1/2}\sum_{\bm R}
e^{-\ii\bk\cdot(\bm R+\bm\tau_\alpha)}c_{\bm R\alpha}$,
where $N_c$ is the number of primitive cells and we set the lattice constant to $1$.
This wavefunction modifies the critical $C=1$ flat band of
Ref.~\cite{Li2026} (reviewed in Section~\ref{supp:critical-start}) by including a weight $\propto \delta$ in an auxiliary orbital.
The components $F_{1,2}$ vanish at $\Gamma$, leading to the critical nature of the original model~\cite{Li2026}. Here, $f(\Gamma)=1$ and so any nonzero $\delta$ removes the gaplessness while preserving $p4$
symmetry.

A finite-range parent Hamiltonian realizing this exactly flat band is
\begin{equation}
 h_\delta(\bk)=E_*\bigl[N_\delta(\bk)I_3
                   -u_\delta(\bk)u_\delta^\dagger(\bk)\bigr],
 \label{eq:gapped-parent}
\end{equation}
where $E_*>0$ sets the overall energy scale.
This $p4$-symmetric Hamiltonian has spectrum
$(0,E_*N_\delta,E_*N_\delta)$ and hopping range $\sqrt5$, measured
between orbital centers. For $\delta\ne0$, $N_\delta$ is everywhere
positive, so the zero-energy flat band is isolated and has $C=0$~\cite{Chen2014}.
The gap at $\Gamma$ is $E_*\delta^2$.

Since a nonzero $\delta$ leads to a flat trivial band, it must also redistribute the Berry curvature. Nevertheless, for small $|\delta|$ the Bloch states remain nearly unchanged outside the $\Gamma$-centered core, and so do the corresponding form factors entering the projected interaction.
In the vicinity of $\Gamma$, however, $\delta$ leads to a compensating negative-curvature core to maintain the overall triviality of the detached flat band.
Such controlled modification of the quantum geometry of the band provides a route to stabilizing an FCI
despite $C=0$; see Sec.~\ref{supp:curvature-redistribution} for the Berry curvature redistribution.
In the following, we fix $\delta=1$ and the corresponding quantum geometry characterization is shown in Figs.~\ref{fig:liyq-model}(b),(c).

\begin{figure}[!t]
 \centering
 \includegraphics[width=\columnwidth]{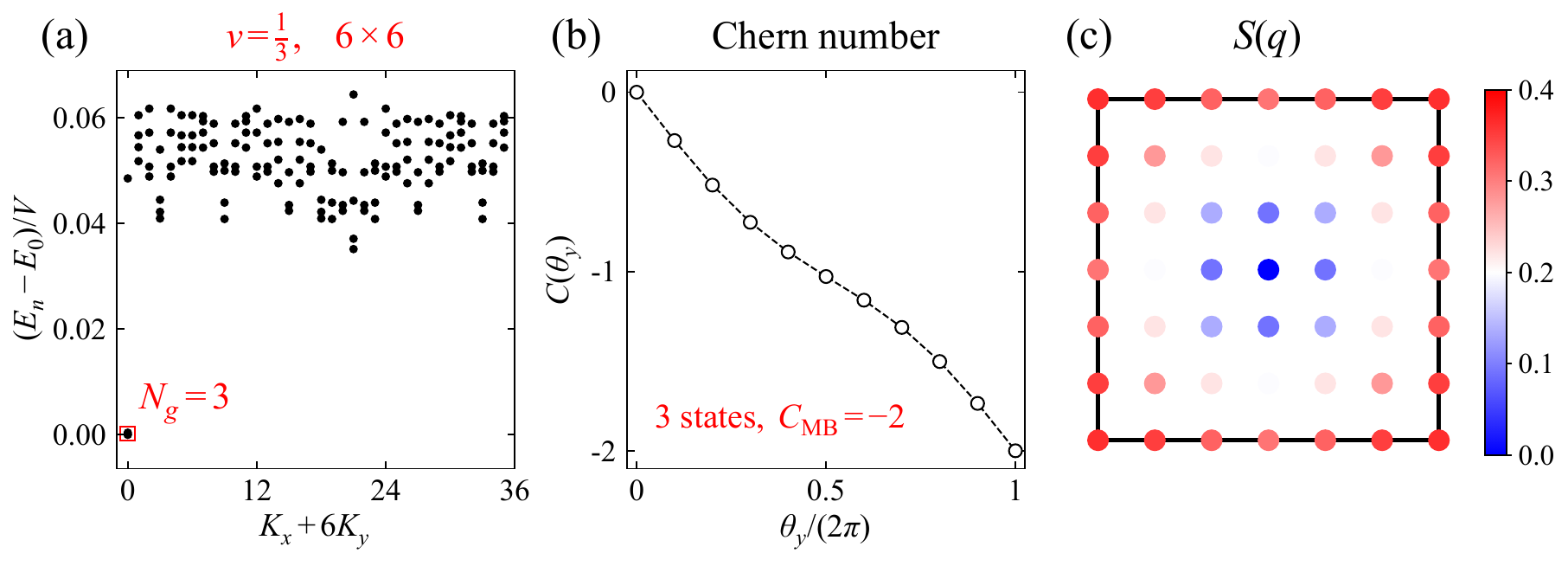}
 \caption{ED of the projected three-band model at one-third filling
on a $6\times6$ torus ($N_e=12$), with $\delta=1$.
(a) Four lowest energy levels in each momentum sector, measured
relative to the ground-state energy and in units of the repulsion $V$.
The red box encloses the three low-energy states at $\bm K=(0,0)$.
(b) Accumulated many-body Berry curvature $C(\theta_y)$ of this
three-state manifold on a $10\times10$ twist grid, yielding
$C(2\pi)=C_{\rm MB}=-2$.
(c) Connected physical-density structure factor $S(\bm q)$,
averaged over the three states and shown over one Brillouin zone.
 }
 \label{fig:one-third-ed}
\end{figure}

To investigate if FCIs emerge, we add nearest-neighbor repulsion between the two $p$ orbitals,
\begin{equation}
 H_V=V\sum_{\bm R}\sum_{\bm d\in\mathcal D}
 n_{\bm R,p_1}n_{\bm R+\bm d,p_2},
 \label{eq:projected-interaction}
\end{equation}
where $\mathcal D=\{0,-\bm e_x,\bm e_y,-\bm e_x+\bm e_y\}$
specifies four symmetry-related bonds of length $1/\sqrt2$.
We then project $H_V$ into the isolated flat band at $\delta=1$.
At one-third filling, ED on a $6\times6$ torus gives three nearly
degenerate ground states at $\bm K=(0,0)$, separated from higher
levels by a sizable many-body gap [Fig.~\ref{fig:one-third-ed}(a)].
To determine the Hall response, we impose boundary twists
$\bm\theta=(\theta_x,\theta_y)$ on the torus. For an $N_g$-state
ground-state manifold, $\sigma_{xy}=(e^2/h)C_{\rm mean}$, where
$C_{\rm mean}=C_{\rm MB}/N_g$ and $C_{\rm MB}$ is the total many-body
Chern number~\cite{Niu1985}, which can be extracted by computing
\begin{equation}
 C(\Theta_y)=\frac{1}{2\pi}
 \int_0^{\Theta_y}\!\dd\theta_y\int_0^{2\pi}\!\dd\theta_x\,
 \mathcal F(\bm\theta),
 \label{eq:main-many-body-chern}
\end{equation}
where $\mathcal F=\ii\sum_{a=1}^{N_g}
[\langle\partial_x\psi_a|\partial_y\psi_a\rangle-\mathrm{c.c.}]$
is the trace Berry curvature of the orthonormal ground states
$|\psi_a(\bm\theta)\rangle$ spanning the manifold, and
$\partial_\mu\equiv\partial/\partial\theta_\mu$.
Figure~\ref{fig:one-third-ed}(b) gives $C_{\rm MB}=C(2\pi)=-2$;
with $N_g=3$, this yields $\sigma_{xy}=-(2/3)e^2/h$.

Aside from the Hall response, we also test for charge ordering using the connected density structure factor
\begin{equation}
 S(\bm q)=\frac{1}{N_c}\bigl[\langle n_{-\bm q}n_{\bm q}\rangle
              -\langle n_{-\bm q}\rangle\langle n_{\bm q}\rangle\bigr],
 \label{eq:main-structure-factor}
\end{equation}
where 
$n_{\bm q}=\sum_{\bm R,\alpha}
e^{\ii\bm q\cdot(\bm R+\bm\tau_\alpha)}n_{\bm R,\alpha}$ includes
all three orbitals. We evaluate the expectation values in the equal-weight ground-state mixture
$\rho_G=\tfrac13\sum_{a=1}^3|\psi_a\rangle\langle\psi_a|$ at zero twist.
The absence of sharp peaks in Fig.~\ref{fig:one-third-ed}(c) rules out
charge order. These results establish an FCI in the isolated $C=0$
band. Section~\ref{supp:three-band-diagnostics} supplements these
results with particle-entanglement spectra and the modular $S$ matrix.
In contrast, at two-thirds filling an integer Hall response
coexists with charge order, forming a quantum Hall crystal, as
detailed in Sec.~\ref{supp:two-thirds}.

\begin{figure}[!t]
 \centering
 \includegraphics[width=\columnwidth]{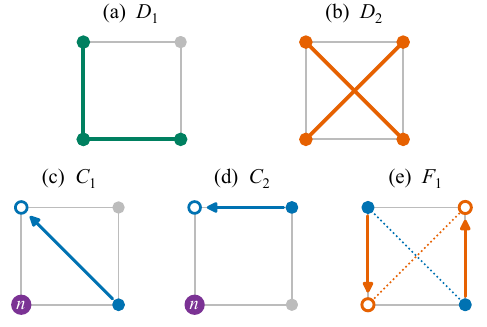}
 \caption{Five classes of local interactions in the plaquette model.
 (a),(b) Nearest-neighbor and diagonal density interactions.
 (c),(d) Representative density-assisted diagonal and edge transfers;
 purple sites mark the density operator, and arrows point from
 annihilation to creation sites. (e) Pair hopping between the two
 plaquette diagonals. Section~\ref{supp:wannier} specifies the full
 Hamiltonian, including translations, Hermitian conjugates, and
 coefficients.
 }
 \label{fig:five-interactions}
\end{figure}

\section{Interaction-only plaquette model}
The emergence of an FCI in a $C=0$ flat band allows us to further take our model to the single-orbital, interaction-only limit by downfolding to the associated Wannier basis. Here, we can directly construct exponentially localized Wannier orbitals from the normalized Bloch state in Eq.~\eqref{eq:regularized-section}~\cite{Brouder2007}, as described in Sec.~\ref{supp:exact-wannier}. Rotating into the Wannier basis leads to a one-band model, with exponentially decaying interaction terms, which is equivalent to the original band-projected model.

To obtain a finite-range model, we truncate the interaction terms within a plaquette. This leads to five interaction terms, labeled as
$D_1,D_2,C_1,C_2,F_1$ in Fig.~\ref{fig:five-interactions}. Each term
acts within one elementary plaquette, with maximum separation $\sqrt2$.
The two density terms $D_{1,2}$ act on nearest-neighbor and diagonal bonds;
the other three describe density-assisted diagonal and edge transfers,
and pair hopping between the diagonals. 
To stabilize the FCI state in the plaquette model, we optimize the finite-size many-body gap by tuning the relative strengths of the five interaction terms while preserving their phase patterns in the truncated interaction parameters.
Summing the same plaquette processes over all
cells preserves all spatial symmetries, giving
\begin{equation}
 H_P=\sum_{\bm R}\sum_{(pq;rs)\in\mathcal L}
 T_{pq;rs}f_p^\dagger f_q^\dagger f_s f_r,
 \label{eq:main-five-interactions}
\end{equation}
where $\mathcal L$ is the set of index tuples $(pq;rs)$ for the retained
plaquette processes in these five classes, including their
off-diagonal Hermitian conjugates. We use $f_a$ as shorthand for
$f_{\bm R+\bm a}$, suppressing the common origin $\bm R$ of each
translated plaquette.
The exact projection and full coefficient table are given in
Sec.~\ref{supp:wannier}.
ED confirms that the FCI survives spatial truncation in the plaquette
model, retaining $\sigma_{xy}=-(2/3)e^2/h$ without charge order. Its
spectrum, many-body Chern number, and density correlations are shown in
Sec.~\ref{supp:one-orbital-ed}.

\begin{figure}[!t]
 \centering
 \includegraphics[width=\columnwidth]{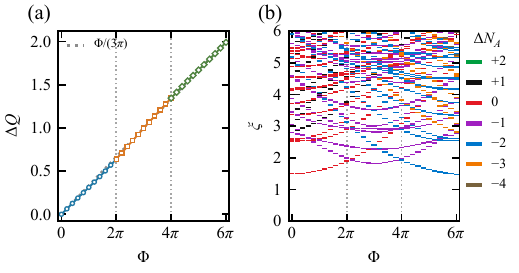}
 \caption{DMRG for the plaquette model on a $24\times8$
 cylinder at filling $1/3$ ($N_e=64$, $D=2000$).
 (a) Cumulative charge leaving the left half of the cylinder. The dotted line indicates a transfer of \(2/3\) of an electron per flux quantum. 
 (b) Charge-resolved entanglement spectrum, cut after column twelve
 (site 96), with $\Delta N_A=N_A-32$. Each dash is one level
 ($\xi\le6$). Both panels use the same continuous $0$--$6\pi$
 trajectory, with $\theta_y=\Phi$ and no fixed-flux reoptimization.
 Vertical dotted lines separate cycles.}
 \label{fig:dmrg-pump-es}
\end{figure}

To investigate whether the FCI survives in the thermodynamic limit,
we further perform DMRG calculations on a $24\times8$ cylinder.
We insert a flux $\Phi$ through the cylinder by imposing twisted
boundary conditions along $y$. We measure the charge leaving the left half,
\begin{equation}
 \Delta Q(\Phi)=\langle N_L\rangle_{\theta_0}
              -\langle N_L\rangle_{\theta_0+\Phi},
 \label{eq:main-pumped-charge}
\end{equation}
where $N_L=\sum_{x<12,y}n_{x,y}$ and $\theta_0=0$.
Figure~\ref{fig:dmrg-pump-es}(a) shows that approximately $2/3$ of an
electron is transferred per $2\pi$ flux, or two electrons over three cycles,
consistent with the magnitude of the Hall response observed in ED.

Figure~\ref{fig:dmrg-pump-es}(b) shows the charge-resolved entanglement
spectrum during flux insertion~\cite{Yang2025}. After $6\pi$ flux
insertion, the low-lying spectral pattern approximately recurs with
$N_A\to N_A-2$, consistent with the transfer of two electrons.
The fractional Hall response in ED and fractional charge pumping in
DMRG establish an FCI generated entirely by single-plaquette
interactions in the plaquette model.

\begin{figure}[!t]
 \centering
 \includegraphics[width=\columnwidth]{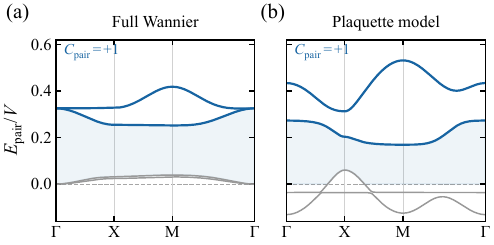}
 \caption{Two-particle interaction spectra of (a) the full Wannier
 interaction before truncation and (b) the plaquette model along
 $\Gamma$--$X$--$M$--$\Gamma$ in total pair momentum $\bm Q$.
 The two strongest repulsive channels (blue) are separated from the
 remaining channels by the shaded gap and carry
 $C_{\mathrm{pair}}=+1$ in both models.}
 \label{fig:pair-topology}
\end{figure}

\section{Two-particle topology and minimality}
\begin{samepage}
With one orbital per cell and no hopping, at the one-particle level the plaquette model is in a strict atomic limit with point-localized orbitals. Such a starting point can be regarded as the extreme opposite of the LL paradigm, in which the single-particle bands mimic the quantum geometry of the LLs.
Nevertheless, our construction reveals that an FCI can still be stabilized solely using finite-range interaction terms.
Intuitively, it is possible for the nontrivial topology of the original multi-band model to remain in the two-particle, correlated sectors. This perspective can be further quantified by examining which two-fermion states
are penalized by the plaquette interaction. At fixed total momentum $\bm Q$, we use the
relative-separation basis
$|\bm Q,\bm d\rangle=N_c^{-1/2}\sum_{\bm R}e^{\ii\bm Q\cdot\bm R}
f_{\bm R}^\dagger f_{\bm R+\bm d}^\dagger|0\rangle$,
with one representative of each pair $\pm\bm d$. 
In this basis, the interaction at fixed total momentum $\bm Q$ is
represented by the matrix~\cite{Lauchli2013Hierarchy}
\begin{equation}
 [H_{\rm eff}(\bm Q)]_{\bm d\bm d'}
 =\langle\bm Q,\bm d|H_{\rm int}|\bm Q,\bm d'\rangle .
 \label{eq:main-pair-heff}
\end{equation}
Its eigenvalues $E_n(\bm Q)$ form the two-particle bands, and its
eigenvectors give the corresponding pair states,
\begin{equation}
 |\Psi_{n\bm Q}\rangle=\sum_{\bm d}
 u_{n,\bm d}(\bm Q)|\bm Q,\bm d\rangle.
 \label{eq:main-pair-wavefunction}
\end{equation}
\par
\end{samepage}

For the full Wannier interaction [Fig.~\ref{fig:pair-topology}(a)],
two strongly repulsive bands lie well above the rest of the spectrum.
Although the relative-separation basis is infinite in the thermodynamic
limit, the four repulsive bonds per unit cell in the three-band model
give rise to only four nonzero two-particle bands after projection
(see Sec.~\ref{supp:full-pair-spectrum} for the derivation).
In the reduced plaquette model [Fig.~\ref{fig:pair-topology}(b)], interactions act only between particles at four relative separations: $\bm d=(0,1),(1,0),(1,-1),(1,1)$.
Truncating and tuning interaction parameters change the dispersion of all four bands but
leave the upper two separated from the rest throughout the Brillouin zone.

The spectra in Fig.~\ref{fig:pair-topology} identify the two strongly
repulsive bands. 
Furthermore, one can define a Chern number through the variation of the eigenvectors over the
center-of-mass Brillouin zone from the projector
$P_2(\bm Q)=\sum_{n=1}^2u_n(\bm Q)u_n^\dagger(\bm Q)$~\cite{Okuma2023Pair,Okuma2025Pair},
\begin{equation}
 C_{\mathrm{pair}}=\frac{\ii}{2\pi}\int_{\mathrm{BZ}}\dd^2Q\,
 \Tr\!\left(P_2[\partial_{Q_x}P_2,\partial_{Q_y}P_2]\right).
 \label{eq:main-pair-chern}
\end{equation}
We find the two-band total Chern number to be $C_{\mathrm{pair}}=+1$ for both the full Wannier interaction
and $H_P$, validating that the reduction from the Wannierized model to the plaquette model preserves the topology of the two-particle bands.

Curiously, a similar two-particle band structure had been observed in an early study of FCIs
on the checkerboard lattice~\cite{Lauchli2013Hierarchy}, which interpreted the two higher-energy two-particle bands as a lattice analog of the $V_1$ Haldane pseudopotential~\cite{Haldane1983}. More recent works similarly identify the gap and topological feature of the two-particle bands as an indicator of potential stabilization of an FCI \cite{Okuma2023Pair, Okuma2025Pair}.
However, in these earlier studies on multi-band models the two-particle bands inherit the nontrivial topology from the single-particle Hamiltonian, similar to how the Haldane pseudopotential gives rise to fractional quantum Hall states only when LL projections are incorporated. In contrast, in our constructed plaquette model, both the topology and energy feature of the two-particle problem arise from the interaction terms.

Furthermore, we find that all of the density-assisted and pair hopping processes $C_{1}$, $C_2$, and $F_1$ are required to preserve a well-defined $C_{\mathrm{pair}}=+1$ for the upper two repulsive bands. In Sec.~\ref{supp:pair-minimality}, we show that removing any one of them 
either closes the pair
gap or makes the isolated pair subspace topologically trivial.
Together, these processes
form the minimal correlated core required to preserve the inherited pair topology within this single-plaquette family.

\section{Conclusion and outlook}
We have constructed a fermionic FCI model with one orbital per unit
cell, no single-particle hopping, and interactions confined to a single
plaquette. Exact diagonalization and DMRG charge pumping on cylinders
support the fractional Hall phase. The topology of its repulsive pair
channels suggests a way to understand and design fractional Hall states
without relying on single-particle band topology or geometry.
An open question concerns the nature of the FCI realized in this work. Unlike the usual lattice analog of the Laughlin state, it exhibits a mismatch between Hall conductance and filling. Whether this mismatch reflects the unusual single-particle starting point remains open.
Beyond the square-lattice example
studied here, it would be interesting to explore which other lattices
admit such interaction-only models and whether they can support
non-Abelian FCIs.

\begin{samepage}
\section*{Acknowledgments}
We thank Chen Fang and Hong-Hao Tu for helpful discussions.
This work is supported by the Hong Kong Research Grants Council through grants AoE/P-604/25-R, C7037-22G, and 16310625, the Croucher Foundation (CIA23SC01), and the Future Science Awards Foundation through an Asian Young Scientist Fellowship. \\

We used ChatGPT Sol 5.6, Sol 6.1, and Astra 6.1 for numerical
simulations, exploration of proof strategies, and manuscript editing.
All code, numerical results, and analytical derivations were reviewed
by the authors.

\section*{Data Availability}

The data that support the findings of this work are available from the authors upon reasonable request.

\par
\end{samepage}

\bibliography{refs}

\onecolumngrid
\clearpage
\section*{Supplementary Material}
\setcounter{section}{0}
\setcounter{equation}{0}
\setcounter{figure}{0}
\renewcommand{\thesection}{S\arabic{section}}
\renewcommand{\theequation}{S\arabic{equation}}
\renewcommand{\thefigure}{S\arabic{figure}}
\renewcommand{\theHsection}{supp.\arabic{section}}
\renewcommand{\theHequation}{supp.\arabic{equation}}
\renewcommand{\theHfigure}{supp.\arabic{figure}}
\renewcommand{\thesubsection}{\Alph{subsection}}
\makeatletter
\renewcommand{\p@subsection}{\thesection\,}
\makeatother
\setcounter{secnumdepth}{2}
\section{Three-band structure and hybridization}
\label{supp:delta}

\subsection{Critical flat-band starting point}
\label{supp:critical-start}

\begin{figure}[!ht]
 \centering
 \includegraphics[width=0.98\textwidth]{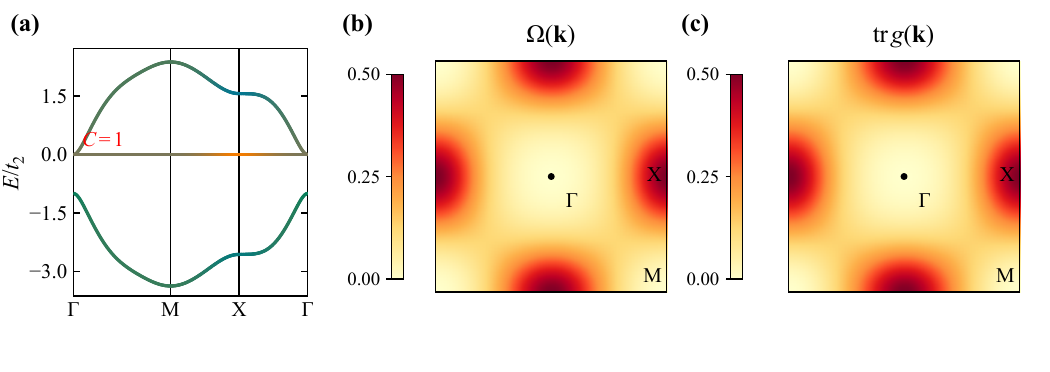}
 \caption{Original critical three-band model at $t_1=2$ and
 $t_2=\Delta=1$.
 (a) Spectrum along $\Gamma$--$M$--$X$--$\Gamma$, colored by the
 orbital probabilities using the blue, orange, and green palette of
 Fig.~\ref{fig:liyq-model}(a). The middle band is exactly flat at zero
 energy, has $C=1$, and touches the upper band at $\Gamma$.
 (b),(c) Berry curvature and metric trace of this critical band in the
 embedding convention. These are the properties before the
 auxiliary component is mixed into the zero-band state.}
 \label{fig:critical-band-structure}
\end{figure}

The construction modifies the fine-tuned square-lattice model of
Ref.~\cite{Li2026}, with the orbital arrangement in
Fig.~\ref{fig:liyq-model}(a). The two $p_-$ orbitals occupy the
$2c$ Wyckoff positions and the auxiliary $s$ orbital occupies $1b$.
We use the embedding convention
$c_{\bk\alpha}=N_c^{-1/2}\sum_{\bm R}
e^{-\ii\bk\cdot(\bm R+\bm\tau_\alpha)}c_{\bm R\alpha}$
and set the lattice constant and $t_2$ to unity. The original
Hamiltonian and its coupling column are
\begin{equation}
 h(\bk)=
 \begin{pmatrix}0_{2\times2}&S(\bk)\\S^\dagger(\bk)&-\Delta\end{pmatrix},
 \qquad \Delta>0,
 \label{eq:liyq}
\end{equation}
\begin{equation}
 S(\bk)=\begin{pmatrix}
 t_1\sin(k_x/2)-\ii\cos(k_x/2)\sin k_y\\
 \cos(k_y/2)\sin k_x-\ii t_1\sin(k_y/2)
 \end{pmatrix}.
 \label{eq:coupling}
\end{equation}
Its energies are
\begin{equation}
 E_0=0,\qquad E_\pm=\frac{-\Delta\pm\sqrt{\Delta^2+4q}}{2},
 \qquad q(\bk)=S^\dagger(\bk)S(\bk).
 \label{eq:supp-original-bands}
\end{equation}
Its zero-band Bloch vector is $u_0=(F^T,0)^T$, with
$F=(-S_2^*,S_1^*)^T$ and $F^\dagger F=q$.
The zero band has no auxiliary-orbital weight, even though the other
two bands mix the physical and auxiliary orbitals.
At $t_1=2$ and $t_2=1$, the coupling near $\Gamma$ is
$S=(k_x-\ii k_y)(1,1)^T+O(|\bk|^3)$.
Thus $q=2|\bk|^2+O(|\bk|^4)$ and
$E_+=2|\bk|^2/\Delta+O(|\bk|^4)$, giving a quadratic touching.
Although $F$ vanishes at $\Gamma$, its projector has the unique limit
$P_0(\Gamma)=|p\rangle\langle p|$, with
$|p\rangle=(-1,1,0)^T/\sqrt2$.
Figure~\ref{fig:critical-band-structure} shows the three-band spectrum
and the quantum geometry of the critical $C=1$ band.
The ratio $t_1=2t_2$ is fine-tuned rather than imposed by $p4$
symmetry. Deviating from it makes the flat band singular, so the
Berry curvature at $\Gamma$ and the band Chern number become ill
defined.

\subsection{Isolating the hybridized flat band}
\label{supp:isolated-band}

Hybridization in Eq.~\eqref{eq:regularized-section} adds the auxiliary
component $\delta f$. At $t_1=2$ and $t_2=1$, $F$ vanishes only at
$\Gamma$, whereas $f(\Gamma)=1$.
The zeros of $f$ lie on the zone boundary, where $F\ne0$.
Thus the squared norm
$N_\delta(\bk)\equiv u_\delta^\dagger(\bk)u_\delta(\bk)
=F^\dagger F+\delta^2|f|^2$ is strictly positive for $\delta\ne0$.
The flat-band projector is
\begin{equation}
 P_\delta(\bk)=\frac{u_\delta(\bk)u_\delta^\dagger(\bk)}{N_\delta(\bk)}.
 \label{eq:projector}
\end{equation}
The two complementary bands of Eq.~\eqref{eq:gapped-parent} have energy
$E_*N_\delta$, so the same condition opens a gap above the zero band.
This parent is reconstructed from the zero-band state, rather than
obtained by adding a term to the original Hamiltonian. In particular,
$h_{\delta=0}=E_*\operatorname{diag}(SS^\dagger,q)$ shares the
critical zero-band projector with $h$ but not its full spectrum.
The auxiliary-orbital probability in the normalized band state is
$w_s(\bk)\equiv\langle s|P_\delta(\bk)|s\rangle
=\delta^2|f(\bk)|^2/N_\delta(\bk)$, which equals one at $\Gamma$.

Under a fourfold rotation,
$u_\delta(R_4\bk)=U_4u_\delta(\bk)$, where
$R_4(k_x,k_y)=(-k_y,k_x)$ and $U_4=(-\ii\sigma_x)\oplus1$.
The orbital positions determine its signs across reciprocal boundaries.
The Bloch frame
$v_{\delta,\alpha}(\bk)=e^{\ii\bk\cdot\bm\tau_\alpha}
u_{\delta,\alpha}(\bk)/\sqrt{N_\delta(\bk)}$
is periodic and real analytic, which establishes $C=0$.

\subsection{Hybridization-induced curvature redistribution}
\label{supp:curvature-redistribution}

\begin{figure}[!ht]
 \centering
 \includegraphics[width=0.68\textwidth]{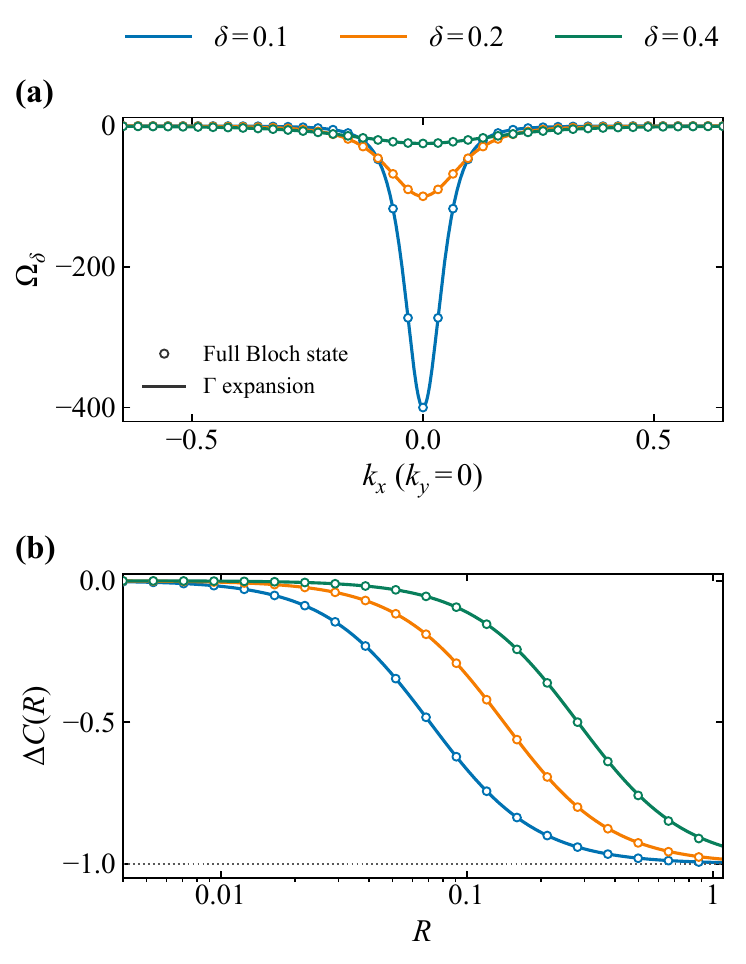}
 \caption{Hybridization-induced curvature redistribution for
 $\delta=0.1$, $0.2$, and $0.4$.
 Open circles use the full Bloch state; solid curves use
 the local expansion in Eqs.~\eqref{eq:local-spinor}
 and \eqref{eq:curvature-core}.
 (a) Berry curvature along $k_y=0$, compared with
 $\Omega_0+\Omega_{\rm core}$.
 (b) Excess disk flux
 $\Delta C(R)=(2\pi)^{-1}\int_{|\bk|<R}\Delta\Omega\,\dd^2k$,
 compared with the local disk integral in Eq.~\eqref{eq:curvature-core}.
 The dotted line marks $-1$.}
 \label{fig:delta-transfer}
\end{figure}

To understand how hybridization changes the quantum geometry, we
derive the Berry curvature near the former touching. We use the
embedding convention of the main text, with
$\mathcal A_i=\ii\langle\psi|\partial_i\psi\rangle$ and
$\Omega=\partial_x\mathcal A_y-\partial_y\mathcal A_x$.
In terms of the unnormalized Bloch vector,
\begin{equation}
 \Omega_\delta=-2\operatorname{Im}\biggl[
 \frac{\langle\partial_xu_\delta|\partial_yu_\delta\rangle}{N_\delta}
 -\frac{\langle\partial_xu_\delta|u_\delta\rangle
 \langle u_\delta|\partial_yu_\delta\rangle}{N_\delta^2}\biggr].
 \label{eq:unnormalized-curvature}
\end{equation}
We denote the original critical-band curvature by $\Omega_0$.
Near $\Gamma$, let $z=k_x+\ii k_y$ and $r=|z|$. Expanding gives
$F=z(-1,1)^T+O(r^3)$ and $f=1-r^2/8+O(r^4)$.
Substituting into $u_\delta$ and normalizing yields
\begin{equation}
 |\psi_\delta\rangle\simeq
 \frac{\sqrt{2}z|p\rangle+\delta|s\rangle}
      {\sqrt{2r^2+\delta^2}},
 \qquad |p\rangle=\frac{(-1,1,0)^T}{\sqrt{2}}.
 \label{eq:local-spinor}
\end{equation}
Here $|s\rangle=(0,0,1)^T$ denotes the auxiliary orbital.
The physical component winds once in phase around $\Gamma$,
while the auxiliary component remains nonzero at the center.
With $z=r e^{\ii\phi}$ and $w(r)=2r^2/(2r^2+\delta^2)$,
Eq.~\eqref{eq:local-spinor} gives the Berry connection
$\mathcal A=-w(r)\dd\phi$, so
$\Omega_{\rm core}=-(1/r)\partial_r w$.
Integrating over a disk yields
\begin{equation}
 \Omega_{\rm core}(r;\delta)
 =-\frac{4\delta^2}{(\delta^2+2r^2)^2},\qquad
 \frac{1}{2\pi}\int_{r<R}\Omega_{\rm core}\,\dd^2k
 =-\frac{2R^2}{\delta^2+2R^2}.
 \label{eq:curvature-core}
\end{equation}
Figure~\ref{fig:delta-transfer} compares the Berry curvature and its
integrated excess flux computed from the full Bloch state with the
local expansion around $\Gamma$. Near $\Gamma$, the curvature correction
$\Delta\Omega\equiv\Omega_\delta-\Omega_0\simeq\Omega_{\rm core}$.
For $|\delta|\ll R\ll1$, its disk flux approaches $-1$, while at fixed
$\bk\ne\Gamma$, Eq.~\eqref{eq:unnormalized-curvature} gives
$\Omega_\delta=\Omega_0+O(\delta^2)$. As $\delta\to0$, the negative
curvature concentrates at $\Gamma$ while its integral approaches
$-2\pi$. It therefore becomes a Dirac delta contribution,
\begin{equation}
 \Omega_\delta(\bk)\longrightarrow
 \Omega_{C=1}(\bk)-2\pi\delta^{(2)}(\bk-\Gamma),
 \label{eq:curvature-limit}
\end{equation}
where $\delta^{(2)}$ is the two-dimensional Dirac delta function.
The negative core cancels the original $C=1$ flux, leaving an isolated
$C=0$ band whose geometry remains close to that of the critical band
over most of the Brillouin zone at small $|\delta|$.

For the $\delta=1$ band used in the many-body calculations,
Figs.~\ref{fig:liyq-model}(b),(c) show a negative curvature core
surrounded by positive curvature, with zero total integral.
In the same embedding convention, the metric is
\begin{equation*}
 g_{ij}=\operatorname{Re}\langle\partial_i\psi_\delta|
 (1-P_\delta)|\partial_j\psi_\delta\rangle.
\end{equation*}
At $\Gamma$, $\Omega=-4$ and $\operatorname{tr}g=4$.
\clearpage
\section{Entanglement diagnostics of the three-band FCI}
\label{supp:three-band-diagnostics}

We supplement the main-text ED results with particle-entanglement
spectra and the modular $S$ matrix. Both diagnostics use the full
single-band projection of $H_V$ in Eq.~\eqref{eq:projected-interaction},
with $\delta=1$ and $V$ as the energy unit.

\subsection{Particle-entanglement spectrum}
\label{supp:pes}

\begin{figure}[!ht]
 \centering
 \includegraphics[width=0.80\textwidth]{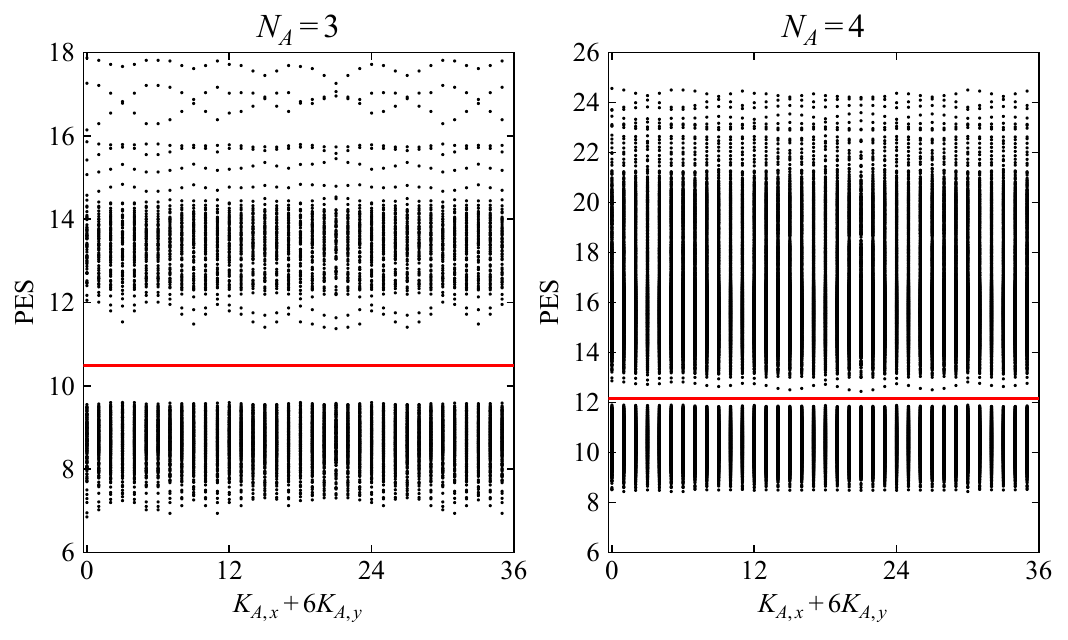}
 \caption{Particle-entanglement spectra of the projected three-band
 model at one-third filling on the $6\times6$ torus, with $\delta=1$
 and $V=1$. Left: $N_A=3$; right: $N_A=4$.
 The horizontal coordinate is $K_{A,x}+6K_{A,y}$.
 Red lines lie inside the entanglement gaps with 4872 and 26325 levels
 below them, respectively.}
 \label{fig:pes}
\end{figure}

We compute the particle-entanglement spectrum (PES) from the
equal-weight mixture of the three zero-twist ground states on the
$6\times6$ torus ($N_e=12$)~\cite{Sterdyniak2011}.
Tracing out $N_B=N_e-N_A$ particles gives
\begin{equation}
 \rho_A=\frac13\sum_{g=1}^{3}\Tr_{N_B}
 |\psi_g\rangle\langle\psi_g|,
 \qquad \xi_j=-\ln\lambda_j,
 \label{eq:pes-density}
\end{equation}
where $\lambda_j$ are the eigenvalues of $\rho_A$, normalized by
$\Tr\rho_A=1$. Translation symmetry resolves the spectrum into
subsystem-momentum blocks $\bm K_A$.
Figure~\ref{fig:pes} shows clear entanglement gaps for $N_A=3$ and $4$,
with 4872 and 26325 levels below the gaps, respectively. Both the
total and momentum-resolved counting match the $(1,3)$ Laughlin
quasihole reference~\cite{Bernevig2012}, revealing Laughlin-type
particle correlations.

\subsection{Minimally entangled states and modular \texorpdfstring{$S$}{S} matrix}
\label{supp:modular-s}

We first express the three ground states on the $6\times6$ torus in the
exact Wannier basis, retaining the full projected interaction. This
representation allows us to divide the torus into two cylinders, each
three columns wide. We extract the modular $S$ matrix using minimally
entangled states (MESs) and a quarter rotation~\cite{Zhang2012MES,Zhu2013MES}.
We obtain an orthonormal MES basis
$\{|\Xi_a\rangle\}_{a=1}^3$ by minimizing the sum of their second
R\'enyi entropies within the three-state manifold. The overlap matrix
$\langle\Xi_a|\hat C_4|\Xi_b\rangle$ represents a quarter rotation in
this basis, with $\hat C_4$ acting on the projected Bloch states.

After rephasing the MESs, we obtain
\begin{equation}
 S_{\rm num}\simeq
 \begin{pmatrix}
 0.579 & 0.576 & 0.576\\
 0.576 & -0.290+0.500\ii & -0.290-0.500\ii\\
 0.576 & -0.290-0.500\ii & -0.290+0.500\ii
 \end{pmatrix}.
 \label{eq:supp-numerical-modular-s}
\end{equation}
It closely matches the three-sector Fourier matrix
$S_{ab}=e^{2\pi\ii ab/3}/\sqrt3$, $a,b=0,1,2$. 
This result is consistent with three Abelian anyon sectors and
complements the Hall response and particle-entanglement counting.

\section{Wannier projection and the local interaction-only model}
\label{supp:wannier}

\subsection{Exact Wannier transformation}
\label{supp:exact-wannier}

In the embedding convention of Eq.~\eqref{eq:regularized-section},
the microscopic orbital $i=(\bm r,\alpha)$ lies at $\bm r+\bm\tau_\alpha$.
The amplitude of the $\bm R$-centered Wannier state on $i$ is
\begin{equation}
 w_\alpha(\bm r-\bm R)=\frac{1}{N_c}\sum_{\bk}
 e^{\ii\bk\cdot(\bm r+\bm\tau_\alpha-\bm R)}
 \psi_{\delta,\alpha}(\bk),
 \label{eq:supp-wannier-orbital}
\end{equation}
where $\psi_{\delta,\alpha}(\bk)=
\langle\alpha|\psi_\delta(\bk)\rangle$.
The corresponding creation operator is
$f_{\bm R}^\dagger=\sum_{\bm r,\alpha}
w_\alpha(\bm r-\bm R)c_{\bm r,\alpha}^\dagger$.
Since $\sum_\alpha|\psi_{\delta,\alpha}(\bk)|^2=1$
at every $\bk$, these modes are orthonormal:
$\{f_{\bm R},f_{\bm R'}^\dagger\}=\delta_{\bm R\bm R'}$.
The factor $e^{\ii\bk\cdot\bm\tau_\alpha}\psi_{\delta,\alpha}(\bk)$
is periodic and analytic in $\bk$, so its Fourier coefficients decay
exponentially with $|\bm r-\bm R|$. Within the projected band, the microscopic creation
operator becomes $c_i^\dagger\mapsto\sum_{\bm R}
w_i^*(\bm R)f_{\bm R}^\dagger$, where
$w_i(\bm R)=w_\alpha(\bm r-\bm R)$.

Let $b=(i,j)$ run over all microscopic repulsive bonds, including
their translations, with each bond counted once. Its strength is
$V_b$ ($V_b=V$ for this model). Define the antisymmetrized
Wannier amplitude $A_b(\bm R_1,\bm R_2)=w_i(\bm R_1)w_j(\bm R_2)
-w_i(\bm R_2)w_j(\bm R_1)$. The exact projected interaction is
\begin{equation}
 H_W^{\mathrm{exact}}=\sum_{1<2,\,3<4}K_{12;34}f_1^\dagger f_2^\dagger f_4f_3,
 \qquad K_{12;34}=\sum_b V_b A_b(1,2)^*A_b(3,4),
 \label{eq:supp-wannier-kernel}
\end{equation}
where $1,\ldots,4$ abbreviate the Wannier centers
$\bm R_1,\ldots,\bm R_4$.
The amplitude $A_b$ accounts for the two ways of placing a fermion
pair on the microscopic bond. Repulsive bonds give a
positive-semidefinite kernel. Equation~\eqref{eq:supp-wannier-kernel}
is the exact Wannier representation; spatial truncation is introduced
in the next section.

Projecting the microscopic density repulsion onto overlapping Wannier
orbitals generates density interactions, density-assisted hopping, and
pair hopping involving two, three, and four distinct centers, respectively.

\subsection{Complete plaquette Hamiltonian}

We truncate the exact Wannier interaction to the five classes listed in
Table~\ref{tab:five-interactions} and tune their overall strengths,
keeping the $D_1$ strength and all coefficient phases fixed.
We select parameters that enlarge the gap above the three lowest states
while reducing their internal splitting across boundary twists.
The resulting Hamiltonian
conserves particle number, has one orbital per cell, and contains no
quadratic term.
With one orbital per cell, the single-particle projector is the scalar
$1$ and its Berry curvature vanishes. The geometry and chirality
inherited from the parent band reside in the quartic amplitudes.

\begin{figure}[!ht]
 \centering
 \includegraphics[width=0.95\textwidth]{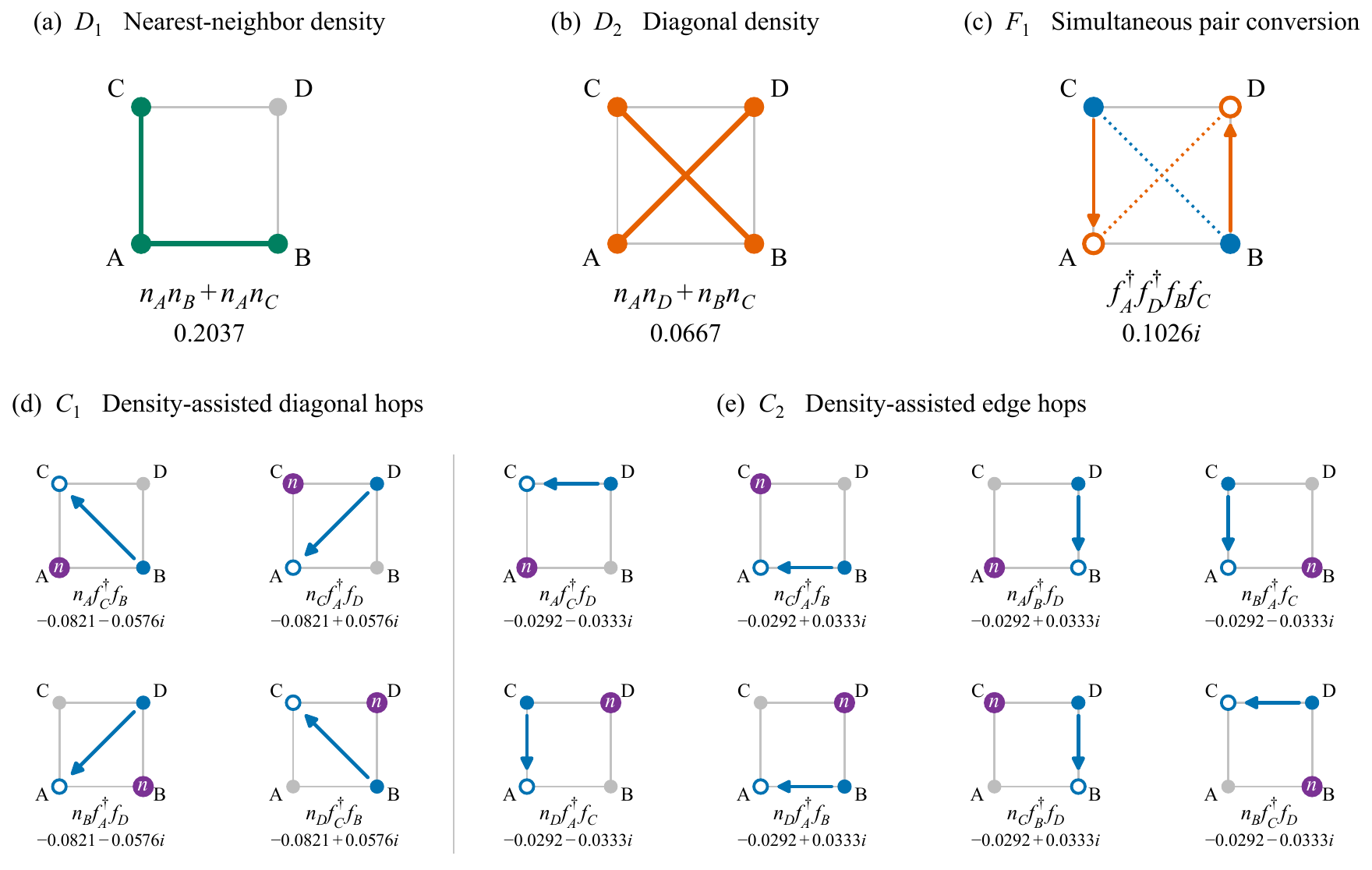}
 \caption{Complete local processes of the five-interaction model.
 Purple circles denote spectator density. Blue arrows in the assisted
 terms point from annihilation to creation sites, with the reordered
 coefficients printed below. The pair-conversion arrows act
 simultaneously. Translations and off-diagonal Hermitian conjugates
 are implicit; no single-particle hopping is present.}
 \label{fig:five-interactions-full}
\end{figure}

All local centers are corners $A=(0,0)$, $B=(1,0)$, $C=(0,1)$ and
$D=(1,1)$ of a square. Table~\ref{tab:five-interactions} specifies the
local terms in Eq.~\eqref{eq:main-five-interactions}, including all
rotation-related processes. We sum the listed terms over all plaquettes
and include the Hermitian conjugate of each hopping term.
Figure~\ref{fig:five-interactions-full} illustrates the density-assisted
hopping processes. Reordering the
operators can change the coefficient's sign relative to its
canonical form in the table.

\begin{table}[!ht]
 \centering
 \begin{tabular}{ccr@{}l}
 \hline\hline
 Class & $(pq;rs)$ & \multicolumn{2}{c}{Coefficient $T_{pq;rs}$}\\
 \hline
 $D_1$ & $(AC;AC)$ & $0.20369994$ & \\
       & $(AB;AB)$ & $0.20369994$ & \\
 $D_2$ & $(AD;AD)$ & $0.06666642$ & \\
       & $(CB;CB)$ & $0.06666642$ & \\
 $C_1$ & $(AC;AB)$ & $-0.08211978$ & $-0.05764975\ii$\\
       & $(AC;CD)$ & $+0.08211978$ & $-0.05764975\ii$\\
       & $(AB;BD)$ & $+0.08211978$ & $+0.05764975\ii$\\
       & $(CD;BD)$ & $-0.08211978$ & $+0.05764975\ii$\\
 $C_2$ & $(AC;AD)$ & $-0.02923914$ & $-0.03332756\ii$\\
       & $(AC;CB)$ & $+0.02923914$ & $-0.03332756\ii$\\
       & $(AB;AD)$ & $-0.02923914$ & $+0.03332756\ii$\\
       & $(AB;CB)$ & $-0.02923914$ & $-0.03332756\ii$\\
       & $(AD;CD)$ & $-0.02923914$ & $-0.03332756\ii$\\
       & $(AD;BD)$ & $-0.02923914$ & $+0.03332756\ii$\\
       & $(CB;CD)$ & $-0.02923914$ & $+0.03332756\ii$\\
       & $(CB;BD)$ & $+0.02923914$ & $+0.03332756\ii$\\
 $F_1$ & $(AD;CB)$ & $0$ & $+0.10256630\ii$\\
 \hline\hline
 \end{tabular}
 \caption{Local coefficients of the tuned five-interaction model,
 in units of the original microscopic $V$. Each entry multiplies
 $f_p^\dagger f_q^\dagger f_s f_r$, with the pair order exactly as listed.
 All translations and off-diagonal Hermitian conjugates are included.}
 \label{tab:five-interactions}
\end{table}

\clearpage
\section{Exact diagonalization of the interaction-only model}
\label{supp:one-orbital-ed}

\begin{figure}[!ht]
 \centering
 \includegraphics[width=0.98\textwidth]{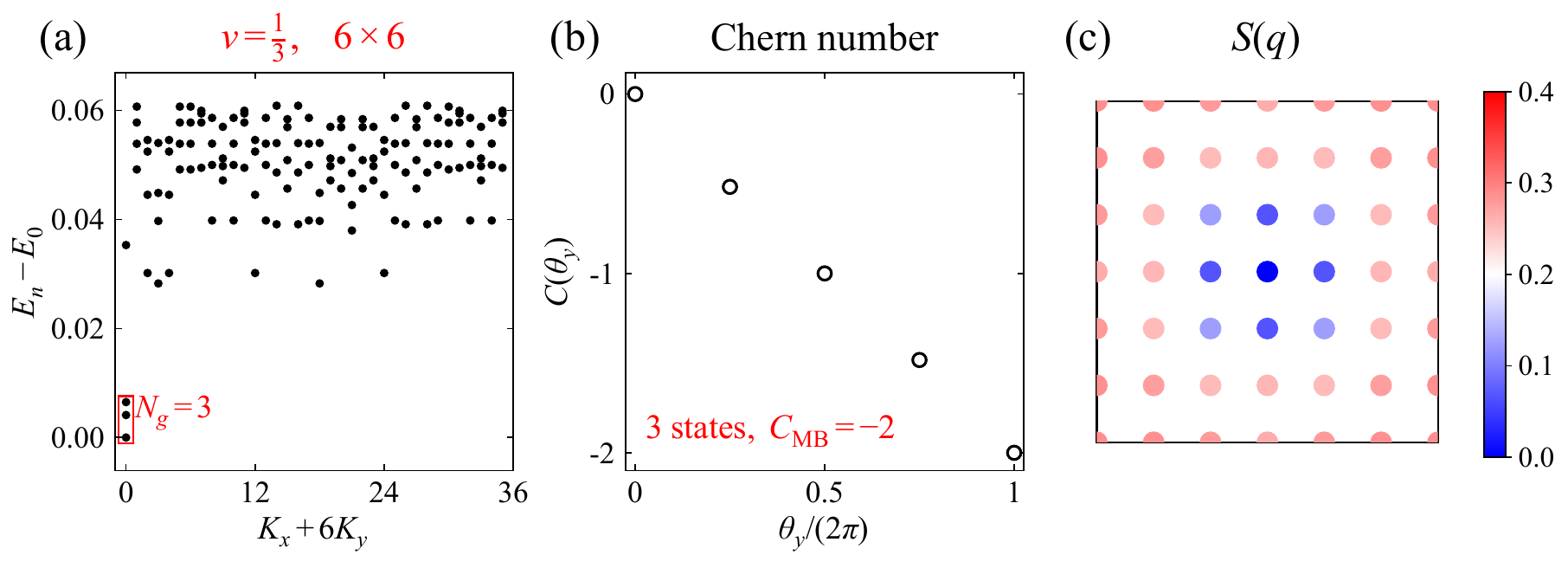}
 \caption{ED of the five-interaction model on a $6\times6$ torus at
 one-third filling. (a) Four lowest levels in every momentum sector;
 the red box marks the three-state manifold at $\bm K=(0,0)$.
 (b) Accumulated Chern number of the rank-three $\bm K=(0,0)$ bundle
 on a $4\times4$ twist mesh. Each circle is a computed cumulative
 value after a $\theta_y$ strip. (c) Connected one-orbital density
 structure factor $S(\bm q)$, averaged over the three states and
 shown over one Brillouin zone.}
 \label{fig:supp-wannier-ed}
\end{figure}

We diagonalize the five-interaction Hamiltonian of
Eq.~\eqref{eq:main-five-interactions} on a $6\times6$ torus at
one-third filling ($N_e=12$). Figure~\ref{fig:supp-wannier-ed}(a)
shows the four lowest levels in each of the 36 momentum sectors.
The three lowest states lie at $\bm K=(0,0)$ and are separated from
the remaining spectrum.

We insert boundary twists through the quartic interaction terms as
described in Sec.~\ref{supp:charge-es}. The rank-three $\bm K=(0,0)$
bundle gives $C_{\rm MB}=-2$ on a $4\times4$ twist mesh
[Fig.~\ref{fig:supp-wannier-ed}(b)]. The plotted points are the
cumulative sums over the four sampled $\theta_y$ strips. At every
sampled twist, the three states remain separated from the fourth
state in this momentum sector, and the link-overlap matrices are
nonsingular. With three states, the mean Chern number is $-2/3$.

For the single orbital, the density operator is
$n_{\bm q}=\sum_{\bm R}e^{\ii\bm q\cdot\bm R}
f_{\bm R}^{\dagger}f_{\bm R}$. We evaluate the connected
$S(\bm q)$ of Eq.~\eqref{eq:main-structure-factor} in the equal-weight
mixture of the three zero-flux states. Figure~\ref{fig:supp-wannier-ed}(c)
shows no sharp charge-order peak. Thus the low-energy spectrum,
fractional Hall response, and density correlations of the
interaction-only Hamiltonian reproduce the corresponding signals
of the projected three-band model.

\section{Flux insertion and charge-resolved Schmidt spectrum}
\label{supp:charge-es}

We thread flux through the cylinder by imposing the twisted boundary
condition $f_{x,y+L_y}=e^{\ii\theta_y}f_{x,y}$.
For an interaction $T_{pq;rs}f_p^\dagger f_q^\dagger f_s f_r$,
the coefficient becomes
$T_{pq;rs}e^{\ii\theta_y(\ell_r+\ell_s-\ell_p-\ell_q)}$,
where $\ell_a$ counts how many times site $a$ crosses the periodic
boundary, with opposite signs for the two directions.
Density interactions are unchanged, while hopping terms acquire a
phase determined by the net charge transferred across the periodic
boundary. The resulting Hamiltonian is $2\pi$ periodic in $\theta_y$.

We insert flux incrementally, initializing each DMRG state from the
preceding one. The charge-pumping trajectory in
Fig.~\ref{fig:dmrg-pump-es}(a) uses a $24\times8$ cylinder with
$N_e=64$ and $D=2000$. Starting at zero flux, we take steps of
$0.2\pi$ with one two-site sweep per step, continuing through three
cycles without fixed-flux reoptimization at the cycle boundaries.
All charge transfers use the initial zero-flux state as a common
reference, as in Eq.~\eqref{eq:main-pumped-charge}. The cumulative
transfers at $2\pi$, $4\pi$, and $6\pi$ are $0.6385$, $1.3444$, and
$1.9904$, respectively.

Figure~\ref{fig:dmrg-pump-es}(b) uses these same 31 states at
$\theta_y=\Phi$ from zero to $6\pi$, with no flux offset or
fixed-flux reoptimization. The cut follows column twelve, or site
96, dividing the 192-site cylinder into equal halves.
For the Schmidt decomposition
$|\psi\rangle=\sum_j\sqrt{p_j}|j_L\rangle|j_R\rangle$, the weights
$p_j$ are the eigenvalues of the left-half reduced density matrix,
and the entanglement energies are $\xi_j=-\ln p_j$.
The reduced density matrix separates into integer $N_A$ blocks.
Colors label $\Delta N_A=N_A-32$, using the same charge origin
throughout the flux insertion. The charge blocks give
$\langle N_A\rangle=\sum_{N_A,j}N_A p_{N_A,j}$, relating the
redistribution of Schmidt weights to charge transport across the cut.
The charge reconstructed from the Schmidt weights agrees with the
direct density measurement to better than $10^{-10}$ at every point.
\clearpage
\section{Two-particle topology and the minimal plaquette mechanism}
\label{supp:pair-topology}

\subsection{Pair channels and their Chern number}

At fixed total momentum $\bm Q$, we expand two-particle states in a
basis of relative displacements,
\begin{equation}
 |\bm Q,\bm d\rangle=\frac{1}{\sqrt{N_c}}
 \sum_{\bm R}e^{\ii\bm Q\cdot\bm R}
 f_{\bm R}^\dagger f_{\bm R+\bm d}^\dagger|0\rangle,
 \qquad
 \bm d\in\{(0,1),(1,0),(1,-1),(1,1)\}.
 \label{eq:supp-pair-basis}
\end{equation}
These four displacements span the subspace on which the plaquette
interaction acts; pairs at other separations have zero interaction
energy. Each term $T_{pq;rs}f_p^\dagger f_q^\dagger f_s f_r$ connects
an incoming pair $(r,s)$ to an outgoing pair $(p,q)$. Reorder these
pairs as $(\bm a_i,\bm a_i+\bm d_i)$ and
$(\bm a_o,\bm a_o+\bm d_o)$ in the basis convention of
Eq.~\eqref{eq:supp-pair-basis}, with fermion signs
$\eta_i,\eta_o=\pm1$. The displacement between the first sites of the two
pairs gives the momentum phase, so
\begin{equation}
 [h_2(\bm Q)]_{\bm d_o\bm d_i}=
 \sum_{(pq;rs)\,\to\,(\bm d_o,\bm d_i)}
 T_{pq;rs}\,\eta_o\eta_i
 e^{\ii\bm Q\cdot(\bm a_i-\bm a_o)},
 \label{eq:supp-pair-record}
\end{equation}
where the sum includes all local terms in
Table~\ref{tab:five-interactions}, including off-diagonal Hermitian
conjugates, that connect the indicated displacements. The translation
sum is already included in Eq.~\eqref{eq:supp-pair-basis}. This gives the
$4\times4$ Hermitian matrix $h_2=H_{\rm eff}$ of
Eq.~\eqref{eq:main-pair-heff}.

Order its eigenvalues as $E_1\ge E_2\ge E_3\ge E_4$.
The two highest-energy states have positive energies and are separated from both
the lower two bands and the inactive zero-energy states. Together
they define the rank-two projector $P_2$ in
Eq.~\eqref{eq:main-pair-chern}. These are the pair configurations
most strongly penalized by the interaction.
This choice follows the logic of Laughlin pseudopotentials: $V_1$
penalizes relative angular momentum $m=1$, while the center-of-mass
states within that sector form a Landau level. The lattice analogue
therefore concerns both the repulsive energies and the
center-of-mass topology of the penalized subspace, measured here by
$C_{\mathrm{pair}}$~\cite{Okuma2025Pair}.

\subsection{Two-particle spectrum of the full Wannier interaction}
\label{supp:full-pair-spectrum}

The four full-Wannier branches in Fig.~\ref{fig:pair-topology}(a)
are obtained by diagonalizing a $4\times4$ Hermitian overlap matrix
$G(\bm Q)$ at each total momentum $\bm Q$. Its rows and columns
label the four microscopic repulsive bond types in
Eq.~\eqref{eq:projected-interaction}.

In the two-electron sector, a single bond repulsion is
$Vn_i n_j=V|ij\rangle\langle ij|$, where
$|ij\rangle=c_i^\dagger c_j^\dagger|0\rangle$. It penalizes only
the configuration with both orbitals occupied and therefore has
rank one. The original interaction has four bond types per
primitive cell, indexed by $b=1,\ldots,4$, with displacements
$\bm d_b\in\mathcal D$. At fixed total momentum, the translations
of each bond form one Bloch state,
\begin{equation}
 |\bm Q,b\rangle=\frac{1}{\sqrt{N_c}}
 \sum_{\bm R}e^{\ii\bm Q\cdot\bm R}
 c_{\bm R,p_1}^\dagger
 c_{\bm R+\bm d_b,p_2}^\dagger|0\rangle.
 \label{eq:supp-microscopic-pair-bond}
\end{equation}
The two-electron interaction in this momentum sector is
$V\sum_b|\bm Q,b\rangle\langle\bm Q,b|$.
Let $\mathcal P$ project both electrons into the isolated flat band.
The full projected interaction is
\begin{equation}
 H_2^{\mathrm{full}}(\bm Q)
 =V\sum_{b=1}^{4}\mathcal P|\bm Q,b\rangle
                  \langle\bm Q,b|\mathcal P.
 \label{eq:supp-full-pair-rank}
\end{equation}
Each projected bond state can extend over many relative
separations, but projection does not increase the number of
vectors. Thus $H_2^{\mathrm{full}}(\bm Q)$ has rank at most four.
All pair states orthogonal to these four vectors have zero
interaction energy.

The projected bond states span the nonzero-energy subspace but are
generally nonorthogonal. Their overlap matrix is
\begin{equation}
 G_{bb'}(\bm Q)=\langle\bm Q,b|\mathcal P|\bm Q,b'\rangle,
 \qquad b,b'=1,\ldots,4,
 \label{eq:supp-full-pair-overlap-definition}
\end{equation}
where we used $\mathcal P^\dagger\mathcal P=\mathcal P$.
Expand an eigenstate as
$|\Psi\rangle=\sum_b c_b\mathcal P|\bm Q,b\rangle$.
Because these states are not orthogonal, the coefficients $c_b$
satisfy a generalized eigenvalue problem. The overlap matrix is
$G$, while Eq.~\eqref{eq:supp-full-pair-rank} gives the Hamiltonian
matrix elements
\begin{equation}
 \langle\bm Q,b|\mathcal P H_2^{\mathrm{full}}\mathcal P
 |\bm Q,b'\rangle=V(G^2)_{bb'}.
 \label{eq:supp-full-pair-hamiltonian-matrix}
\end{equation}
The generalized eigenvalue equation is therefore
\begin{equation}
 VG^2c=EGc.
 \label{eq:supp-full-pair-reduced-eigenproblem}
\end{equation}
On the nonzero-eigenvalue subspace of $G$, we can cancel one factor
of $G$ to obtain the ordinary eigenvalue equation $Gc=(E/V)c$.
Thus the eigenvalues of the overlap matrix give the pair energies
in units of $V$. The corresponding physical states are normalized
by $c^\dagger Gc=1$.

For the numerical calculation, we absorb the orbital positions into
$v_\alpha(\bk)=e^{\ii\bk\cdot\bm\tau_\alpha}\psi_{\delta,\alpha}(\bk)$.
The bond amplitude is
\begin{equation}
 \mathcal A_b(\bm Q,\bm k)=
 e^{\ii(\bm Q-\bm k)\cdot\bm d_b}
 v_{p_1}(\bm k)v_{p_2}(\bm Q-\bm k)
 -e^{\ii\bm k\cdot\bm d_b}
 v_{p_1}(\bm Q-\bm k)v_{p_2}(\bm k).
 \label{eq:supp-full-pair-amplitude}
\end{equation}
The two terms assign the momenta $\bm k$ and $\bm Q-\bm k$ to the
two orbitals in opposite orders, with the minus sign required by
fermion antisymmetry. The matrix actually diagonalized is
\begin{equation}
 G_{bb'}(\bm Q)=\frac{1}{2N_k}\sum_{\bm k}
 \mathcal A_b(\bm Q,\bm k)\mathcal A_{b'}(\bm Q,\bm k)^*,
 \label{eq:supp-full-pair-gram}
\end{equation}
where $N_k$ is the number of internal momentum points. The factor
$1/2$ removes double counting of the pair
$(\bm k,\bm Q-\bm k)$. We use a $64\times64$ internal momentum
grid. The four eigenvalues of $G(\bm Q)$ are the dimensionless pair energies
$E_n(\bm Q)/V$ plotted in Fig.~\ref{fig:pair-topology}(a); all
remaining two-particle energies are zero.

\subsection{General p4 coefficients and the pair-sector matrix}

To determine which correlated processes are needed for this pair
topology, we allow all $p4$-symmetric coefficients within the five
retained classes. The first two basis states in
Eq.~\eqref{eq:supp-pair-basis} are nearest-neighbor pairs, and the
last two are diagonal pairs. In these two sectors, the pair
Hamiltonian has the block form
\begin{equation}
 h_2(\bm Q)=\begin{pmatrix}A(\bm Q)&K(\bm Q)\\
                          K^\dagger(\bm Q)&B(\bm Q)\end{pmatrix},
 \quad
 A=\begin{pmatrix}d_1&g\\g^*&d_1\end{pmatrix},
 \quad
 B=\begin{pmatrix}d_2&-\ii t_F e^{-\ii Q_y}\\
                  \ii t_F e^{\ii Q_y}&d_2\end{pmatrix},
 \label{eq:supp-pair-blocks}
\end{equation}
where $d_1,d_2,a,b,t_F$ are real and $z_1,z_2$ below are complex.
The explicit momentum dependence is
\begin{align}
 g(\bm Q)&=e^{\ii(Q_y-Q_x)/2}
 \left[-4a\sin\frac{Q_x}{2}\sin\frac{Q_y}{2}
       -4\ii b\cos\frac{Q_x}{2}\cos\frac{Q_y}{2}\right],
 \label{eq:supp-pair-g}\\
 K(\bm Q)&=\begin{pmatrix}
 z_2e^{\ii Q_y}(1+e^{-\ii Q_x})&z_1(1+e^{-\ii Q_x})\\
 z_1(1+e^{\ii Q_y})&-z_2(1+e^{-\ii Q_y})
 \end{pmatrix}.
 \label{eq:supp-pair-K}
\end{align}
Here the class labels also denote coupling amplitudes:
$D_1,D_2$ are the nearest-neighbor and diagonal density strengths.
For the correlated terms, we denote the amplitudes of
the representative processes $(AC;AB)$, $(AC;AD)$, $(AC;CB)$, and
$(AD;CB)$ by $C_1$, $C_{2a}$, $C_{2b}$, and $\ii F_1$,
respectively, using the plaquette labels and operator ordering of
Table~\ref{tab:five-interactions}. The extra subscripts in
$C_{2a},C_{2b}$ distinguish two symmetry-inequivalent processes
within the same $C_2$ class. Their complex amplitudes are
independent in the general $p4$ family, whereas the tuned model
fixes their ratio and rescales them together. We write the
purely imaginary $F_1$ amplitude as $\ii F_1$, with real $F_1$.
The matrix parameters are then
\begin{align}
 d_1&=D_1,\qquad d_2=D_2,\nonumber\\
 a&=-\operatorname{Re}C_1,\qquad
 b=-\operatorname{Im}C_1,\nonumber\\
 z_1&=C_{2a},\qquad z_2=C_{2b},\qquad t_F=F_1.
 \label{eq:supp-pair-coefficients}
\end{align}
The density interactions $D_1$ and $D_2$ therefore contribute
$d_1I_2$ to the nearest-neighbor pair block and $d_2I_2$ to the
diagonal pair block, respectively. The $C_1$ terms determine $g$,
$C_2$ determines $K$, and $F_1$ gives the off-diagonal entries of
$B$. The diagonal pair transfer
$O=f_A^\dagger f_D^\dagger f_B f_C$ transforms as
$O\mapsto-O^\dagger$ under a quarter rotation. Its coefficient must
therefore be purely imaginary. This leaves nine independent real
parameters. The tuned model in Table~\ref{tab:five-interactions}
varies only five real class strengths, keeping all internal
coefficient ratios fixed.

For the coefficients in Table~\ref{tab:five-interactions},
$(d_1,d_2)=(0.20369994,0.06666642)$,
$(a,b)=(0.08211978,0.05764975)$,
$t_F=0.10256630$, $z_1=-0.02923914-0.03332756\ii$, and
$z_2=0.02923914-0.03332756\ii$.
The constraints are $K(M)=0$ and $g(X)=0$,
with $M=(\pi,\pi)$ and $X=(\pi,0)$.

\subsection{Why each correlated class is required}
\label{supp:pair-minimality}

The proofs use a common observation. Suppose the two block spectra
are separated throughout the Brillouin zone, with
$\lambda_{\min}[A(\bm Q)]>\lambda_{\max}[B(\bm Q)]$ at every
$\bm Q$. The Courant--Fischer theorem~\cite{HornJohnson2013}
applied to the two block subspaces gives
\begin{equation}
 E_2\ge\lambda_{\min}(A),\qquad
 E_3\le\lambda_{\max}(B),\qquad
 E_2-E_3\ge\lambda_{\min}(A)-\lambda_{\max}(B)>0.
 \label{eq:supp-pair-minmax}
\end{equation}
These bounds hold for any coupling $K$. Reducing $K$ continuously
to zero therefore keeps the upper two bands separated from the
lower two everywhere. At the endpoint, their projector is
$\operatorname{diag}(I_2,0)$, independent of $\bm Q$, so
$C_{\mathrm{pair}}=0$. The same argument applies with $A$ and $B$
exchanged. Below we write $\mu=d_1-d_2$.

\textit{Removing $C_1$.---}Now $A=d_1I_2$. At $M$, the pair energies
are $d_1,d_1,d_2+|t_F|,d_2-|t_F|$. If
$|\mu|\le|t_F|$, the $d_1$ doublet closes the gap between the second
and third levels. Isolation therefore requires $|\mu|>|t_F|$.
Both block spectra are momentum independent, so this condition
separates them throughout the Brillouin zone. The continuous
decoupling $K\to0$ then gives $C_{\mathrm{pair}}=0$.

\textit{Removing $C_2$.---}Now $K=0$, so the two blocks do not mix.
As $\bm Q$ varies, changing the number of $A$-block states among
the upper two bands requires an exchange with a state in the lower
two bands, closing the gap $E_2-E_3$. This number must therefore
remain fixed throughout the Brillouin zone if the gap stays open.
Since $A(X)=d_1I_2$, it is either zero or two: selecting only one
state would split the degenerate $X$-point doublet between the upper
and lower groups. The upper two states thus span either the entire
$A$ subspace or the entire $B$ subspace at every $\bm Q$.
Their projector is therefore momentum independent, giving
$C_{\mathrm{pair}}=0$.

\textit{Removing $F_1$.---}Now $B=d_2I_2$. At $M$, $K(M)=0$ and
$g(M)=-4a$, so the four energies are
$d_1-4|a|,d_1+4|a|,d_2,d_2$. If $|\mu|\le4|a|$, the $d_2$
doublet occupies the middle two levels, giving $E_2(M)=E_3(M)=d_2$.
An isolated upper-two-band subspace therefore requires
$|\mu|>4|a|$. To extend this local separation to the full Brillouin
zone, we first deform $b$ to zero while preserving the gap.
The coupling block satisfies
\begin{equation}
 \det K=-(z_1^2+z_2^2)
 (1+e^{-\ii Q_x})(1+e^{\ii Q_y}).
 \label{eq:supp-pair-detK}
\end{equation}
For $z_1^2+z_2^2\ne0$, $K$ is invertible away from the boundary lines
$Q_x=\pi$ or $Q_y=\pi$. A zero mode $(v,w)$ of $h_2-d_2I_4$ would
satisfy
\begin{equation}
 (A-d_2I_2)v+Kw=0,\qquad K^\dagger v=0.
 \label{eq:supp-pair-zero-mode}
\end{equation}
Invertibility of $K$ first forces $v=0$ and then $w=0$, so $d_2$
cannot be an eigenvalue for any $A$. At each fixed interior
momentum, the number of eigenvalues above or below $d_2$ therefore
cannot change as $A$ varies. We count them by setting $A=d_2I_2$:
the energies relative to $d_2$ are the two positive singular values
of $K$ and their negatives. Hence there are exactly two levels on
each side, and $E_2>d_2>E_3$ holds for any $A$. In particular,
varying the $C_1$ parameter $b$ cannot close the gap in the interior.

On the boundary lines, $\cos(Q_x/2)\cos(Q_y/2)=0$, so the
$b$-dependent term in $g$ vanishes. Changing $b$ leaves the entire
Hamiltonian unchanged there. Thus the deformation $b\to0$ preserves
the gap throughout the Brillouin zone. At $b=0$,
$|g(\bm Q)|=4|a\sin(Q_x/2)\sin(Q_y/2)|\le4|a|$.
The eigenvalues of $A$ lie in $[d_1-4|a|,d_1+4|a|]$, while both
eigenvalues of $B$ equal $d_2$. The condition $|\mu|>4|a|$ places
$d_2$ outside this interval, separating the block spectra everywhere.
Equation~\eqref{eq:supp-pair-minmax}, with $A$ and $B$ exchanged
when needed, then bounds the gap by
$E_2-E_3\ge|\mu|-4|a|>0$ as $K$ is turned off. At the endpoint,
the upper two states span one complete block with a constant
projector, so $C_{\mathrm{pair}}=0$.

If $z_1^2+z_2^2=0$, we first make a sufficiently small
$p4$-symmetric change of $z_1,z_2$ so that the sum is nonzero.
The initially open gap remains open under a sufficiently small
perturbation, so the total Chern number of the upper two bands is unchanged. The preceding
deformation then applies.

Removing any one of $C_1,C_2,F_1$ therefore either closes the gap
below the upper two bands or makes their Chern number zero, for any
remaining $p4$-symmetric coefficients in this plaquette family.
These three correlated processes form the minimal core of the
inherited pair topology.

\clearpage
\section{Quantum Hall crystal at two-thirds filling}
\label{supp:two-thirds}
\begin{figure}[!ht]
 \centering
 \includegraphics[width=0.95\textwidth]{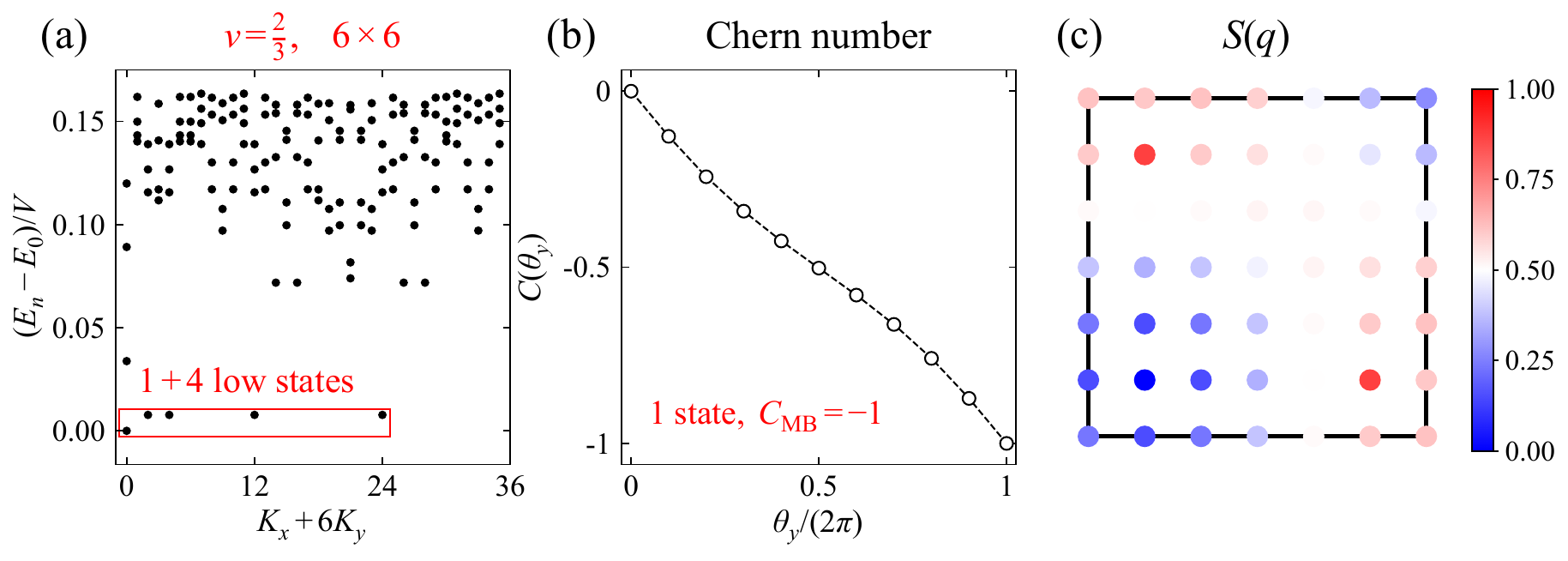}
 \caption{Two-thirds filling of the full projected interaction on a
 $6\times6$ torus ($N_e=24$), with the same parameters as
 Fig.~\ref{fig:one-third-ed}.
 (a) Four lowest levels in each momentum sector. The red box marks
 the ground state and four low-lying states.
 (b) Cumulative Berry flux of the lowest $\bm K=(0,0)$ state on a
 $10\times10$ twist grid, yielding $C_{\rm MB}=-1$.
 (c) Connected physical-density structure factor $S(\bm q)$ of the
 same zero-twist state. The momentum window $q_x,q_y\in[-\pi/3,5\pi/3]$
 places both peaks, $(4\pi/3,0)$ and $(0,4\pi/3)$, in the interior.}
 \label{fig:two-thirds-ed}
\end{figure}

At two-thirds filling ($N_e=24$), the full single-band projection of
$H_V$ on the $6\times6$ torus instead favors charge order. We use
$\delta=1$ and $V$ as the energy unit, and the integer momentum labels
$\bm K$ correspond to $2\pi\bm K/6$.
The ground state at
$\bm K=(0,0)$ is accompanied by four low-lying states at
$(2,0)$, $(4,0)$, $(0,2)$, and $(0,4)$, separated from higher levels
by a many-body gap [Fig.~\ref{fig:two-thirds-ed}(a)]. The density
structure factor exhibits sharp peaks at $(4\pi/3,0)$ and
$(0,4\pi/3)$ [Fig.~\ref{fig:two-thirds-ed}(c)]. These wavevectors
match the momenta of the low-energy states and indicate period-three
charge order.

Following the lowest $\bm K=(0,0)$ state over a $10\times10$ twist
grid gives $C_{\rm MB}=-1$
[Fig.~\ref{fig:two-thirds-ed}(b)],
corresponding to $\sigma_{xy}=-e^2/h$. Together, the integer
Hall response and charge order indicate a quantum Hall crystal.
The contrast with the FCI at one-third filling parallels the
filling asymmetry reported in Ref.~\cite{Niu2025}.

\end{document}